\documentclass[superscriptaddress,aps,preprint]{revtex4}
\usepackage[dvips]{graphicx}

\newcommand{\be}{\begin{equation}}
\newcommand{\en}{\end{equation}}
\newcommand{\ba}{\begin{eqnarray}}
\newcommand{\ea}{\end{eqnarray}}

\newcommand{\Slash}[1]{{#1}\!\!\!/}
\newcommand{\RM}[1]{\mathrm{#1}}

\begin{document}
\title{Dynamical Lorentz symmetry breaking in a $4$D four-fermion model at finite temperature}

\author{M. Gomes}
\affiliation{Instituto de F\'{\i}sica, Universidade de S\~ao Paulo\\
Caixa Postal 66318, 05315-970, S\~ao Paulo, SP, Brazil}
\email{mgomes,ajsilva@fma.if.usp.br}
\author{T. Mariz}
\affiliation{UAST, Universidade Federal Rural de Pernambuco\\
Caixa Postal 063, CEP 56900-000, Serra Talhada, PE, Brazil}
\email{tmariz@uast.ufrpe.br}
\author{J. R. Nascimento}
\affiliation{Departamento de F\'{\i}sica, Universidade Federal da Para\'{\i}ba\\
Caixa Postal 5008, 58051-970, Jo\~ao Pessoa, Para\'{\i}ba, Brazil}
\email{alesandro,jroberto,petrov@fisica.ufpb.br}
\author{A. F. Santos}
\affiliation{Departamento de F\'{\i}sica, Universidade Federal da Para\'{\i}ba\\
Caixa Postal 5008, 58051-970, Jo\~ao Pessoa, Para\'{\i}ba, Brazil}
\email{alesandro,jroberto,petrov@fisica.ufpb.br}
\author{A. Yu. Petrov}
\affiliation{Departamento de F\'{\i}sica, Universidade Federal da Para\'{\i}ba\\
Caixa Postal 5008, 58051-970, Jo\~ao Pessoa, Para\'{\i}ba, Brazil}
\email{alesandro,jroberto,petrov@fisica.ufpb.br}
\author{A. J. da Silva}
\affiliation{Instituto de F\'{\i}sica, Universidade de S\~ao Paulo\\
Caixa Postal 66318, 05315-970, S\~ao Paulo, SP, Brazil}
\email{mgomes,ajsilva@fma.if.usp.br}

\begin{abstract}
In a $4$D four-fermion model we study the dynamical restoration of Lorentz and CPT symmetries at  finite temperature. 
We evaluate the gap equation both at zero and at finite temperature and observe that, depending on the parameters of the theory,
there is a critical temperature at which the Lorentz and CPT symmetries 
are restored.
\end{abstract}


\maketitle

\section{Introduction}

The Lorentz and CPT symmetries, are certainly among the more  important symmetries of nature  (see f.e. \cite{Kostel}).  
However, motivated by the fact that spontaneous breaking of Lorentz symmetry may arise in the context of string theory \cite{Kos01}, the possibility of  Lorentz symmetry breaking is now being intensively discussed for a large class of field models.
In this context, some high precision experiments to measure the scale of this possible  breaking have been proposed \cite{Kost}.   
In the framework of field theory this phenomenon has been studied in Ref.~\cite{Jac}. In Ref.~\cite{Kos01} the standard model 
extension (SME) including all possible Lorentz and/or CPT breaking terms was discussed. Some relevant topics on the Lorentz symmetry breaking can be also found in \cite{Bazeia,Kos02,Blu01,Coll}.

The simplest example of a theory with spontaneous Lorentz symmetry 
breaking is a bumblebee model in which a vector field acquires a nonzero vacuum expectation value (VEV) implying that the vacuum of the theory gets a preferential direction in the spacetime. Different aspects of  bumblebee 
models have been analyzed in the Refs.~\cite{Kos03,Alts,Bert}.

Many versions of the bumblebee models may be defined with different forms of the potential and kinetic terms for the vector field and different
couplings to matter and gravity. Some studies of the coupling of the bumblebee models in a curved space are presented in \cite{Blu02,Blu03,Blu04,Blu05} where this coupling was considered in various spacetime geometries, such as Minkowski, Riemann, or Riemann-Cartan spacetimes. 

Let us introduce our bumblebee model. The generic Lagrangian for it looks like \cite{Blu06}
\begin{eqnarray}
{\cal L}={\cal L}_B+{\cal L}_G+{\cal L}_M,
\end{eqnarray} 
where ${\cal L}_G$ is the pure gravitational term, ${\cal L}_M$ is the matter term and  
\begin{eqnarray}
{\cal L}_B=\sqrt{-g}\Biggl(-\frac{1}{4}B_{\mu\nu}B^{\mu\nu}-V(B_{\mu}, b^2)+B_{\mu}J^{\mu}\Biggl)
\end{eqnarray}
is the bumblebee-like term. Here $ B_{\mu} $ is a vector field, the stress tensor $B_{\mu\nu}$ is defined as
\begin{eqnarray}
B_{\mu\nu}=D_{\mu}B_{\nu}-D_{\nu}B_{\mu},
\end{eqnarray}
the $J^{\mu}$ is a matter current and $V$ is the potential which has the nontrivial minimum, responsible for the Lorentz symmetry 
violation. The most used forms of the potential are 
\begin{eqnarray}
V=\lambda(B_{\mu}B^{\mu}\mp b^2),
\end{eqnarray}
which is linear in $B_{\mu}B^{\mu}$ and  involves a Lagrange multiplier field $\lambda$ and
\begin{eqnarray}
V= \frac{1}{2} \lambda(B_{\mu}B^{\mu}\mp b^2)^2,
\end{eqnarray}
which is quadratic in $B_{\mu}B^{\mu}$ and the Lagrange multiplier $ \lambda $ may assume a constant value.

In this paper, starting from a four-fermion model,  we study spontaneous Lorentz and CPT symmetry breaking via a bumblebee potential dynamically induced
by radiative corrections at finite temperature.  Using the Matsubara formalism, we will discuss the critical temperature 
at which the Lorentz and CPT symmetries are restored. 
 
This paper is organized as follows. As a first step to introduce the temperature, in Section \ref{NEP} we calculate the effective potential  and the gap equation at zero temperature in a noncovariant way  \cite{Gom}. In Section \ref{LSR} we analyze the gap equation at finite temperature and estimate the critical temperature at which the Lorentz symmetry is restored. The Summary is devoted to the discussion of results.

\section{The Gap Equation at Zero Temperature}\label{NEP}

\begin{figure}[h]
	\centering
		\includegraphics{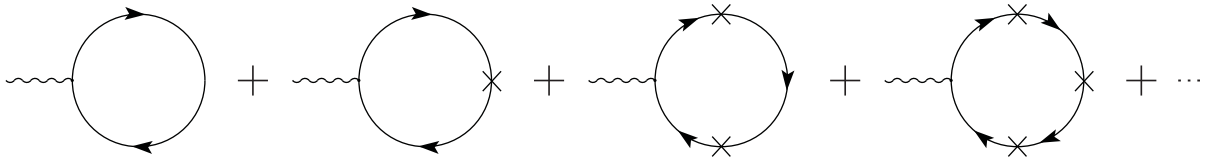}
	\caption{Contributions to the tadpole $\Pi^\mu$}
	\label{fig2}
\end{figure}

Before studying the restoration of Lorentz and CPT symmetries, let us provide more details on the derivation of the nontrivial minima of the effective potential  at zero temperature in a noncovariant way. These details will be useful for the finite temperature calculation of the next section. Our model is described by the Lagrangian
\begin{equation}\label{tThirring}
{\cal L}_0 = \bar\psi(i\Slash{\partial}-m)\psi - \frac G2 (\bar\psi\gamma_\mu\gamma_5\psi)(\bar\psi\gamma^\mu\gamma_5\psi),
\end{equation}
where $ \psi $ is a Dirac spinor field. Following the usual prescription, to eliminate the self-interaction term of the above Lagrangian
we introduce an auxiliary vector field $B_{\mu}$ which leads to an equivalent Lagrangian,
\begin{eqnarray}\label{Lag}
{\cal L} &=& {\cal L}_0 + \frac{1}{2G} \left(B_\mu - {G} \bar\psi\gamma_\mu\gamma_5\psi\right)^2 \nonumber\\
 &=& \frac{1}{2 G} B_\mu B^\mu + \bar\psi(i\Slash{\partial} - m\,-\, \Slash{B}\gamma_5)\psi.
\end{eqnarray}
 Now, by performing the fermionic functional integration we obtain the effective action given by
\begin{equation}\label{1}
S_\RM{eff}[B] = \frac{1}{2 G} \int d^4x\, B_\mu B^\mu -i \RM{Tr} \ln(i\Slash{\partial}-m\,-\,\Slash{B}\gamma_5). 
\end{equation}
The $\RM{Tr}$ stands for the trace over Dirac matrices as well as the for the integration in momentum or coordinate spaces. Thus, the effective potential for a constant $B_{\mu}$ turns out to be
\be\label{Vef}
V_\RM{eff} = -\frac{1}{2G}B_\mu B^\mu + i\,\RM{tr} \int\frac{d^4p}{(2\pi)^4} \, \ln(\,\Slash{p}-m\,-\,\Slash{B}\gamma_5).
\en
To investigate the existence of nontrivial minima, we look for solutions of the expression
\be\label{DVef}
\frac{dV_\RM{eff}}{dB_\mu}\Big|_{B=b} =  - \frac{1}{G} b^\mu - i\,\Pi^\mu = 0,
\en
where
\begin{equation}\label{Pi}
\Pi^\mu = -i\RM{tr} \int \frac{d^4p}{(2\pi)^4} \frac{i}{\,\Slash{p}-m\,-\,\Slash{b} \gamma_5} \gamma^\mu\gamma_5
\end{equation}
is the one-loop tadpole amplitude. To evaluate this integral, we use the same perturbative route applied in \cite{Gom}, where the propagator is the usual $S(p)=i(\Slash{p}-m)^{-1}$, $-i\Slash{b}\gamma_5$ is considered an insertion in this propagator and the trilinear vertex is $-i\gamma^\mu\gamma^5$. With these Feynman rules the contributions to $\Pi^{\mu}$ are shown in Fig.\ref{fig2}. 
By calculating the traces of the Dirac matrices in four dimensions we found that the first and third graphs vanish. The graphs with more than three insertions also vanish \cite{Gom}.

After the calculation of the traces of the Dirac matrices, the second graph of Fig. \ref{fig2} gives
\begin{equation}\label{Pib0}
\Pi_b^\mu = 4i \int \frac{d^4p_\RM{E}}{(2\pi)^4} \frac{(\vec p^2+p_4^2-m^2)b_\RM{E}^\mu-2(p_\RM{E}\cdot b_\RM{E}) p_\RM{E}^\mu}{(\vec p^2+p_4^2+m^2)^2},
\end{equation}
where we have changed the Minkowski space to Euclidean space by performing the Wick rotation $p_0 = ip_4$, $b_0 = ib_4$ and $d^4p = id^4p_\RM{E}$. In order to implement translations only on the spatial components of $p_\RM{E}^\mu=(p_4,\vec p)$, we decompose the $p_\RM{E}^\mu$ as follows \cite{Grig}
\begin{equation}
p_\RM{E}^\mu = \hat p^\mu + p_4\delta^{\mu0},
\end{equation}
where, as a consequence, $\hat p^\mu=(0,\vec p)$. Then, we substitute this expression into Eq. (\ref{Pib0}), yielding
\begin{equation}\label{Pib}
\Pi_b^\mu = \Pi_{1}\, b_\RM{E}^\mu + \Pi_{2}\, b_4\delta^{\mu0},
\end{equation}
with
\begin{equation}
\Pi_{1} = 4i\int\frac{dp_4}{2\pi} (\mu^2)^{\frac{3-D}{2}}\int\frac{d^D\vec p}{(2\pi)^D}\left[\frac{1}{\vec{p}^2+p_4^2+m^2} - \frac{\frac{2}{D}\vec{p}^2+2m^2}{(\vec{p}^2+p_4^2+m^2)^2} \right]
\end{equation}
and
\begin{equation}
\Pi_{2} = -8i\int\frac{dp_4}{2\pi}(\mu^2)^{\frac{3-D}{2}}\int\frac{d^D\vec p}{(2\pi)^D}\left[\frac{1}{\vec{p}^2+p_4^2+m^2} - \frac{\frac{1+D}{D}\vec{p}^2+m^2}{(\vec{p}^2+p_4^2+m^2)^2}\right], 
\end{equation}
where we have promoted the 3-dimensional space integral to  $D$ dimensions, introduced an arbitrary parameter $\mu$ to keep the mass dimension unchanged
 and, due to the symmetry of the integral under spacial rotations,  replaced 
\begin{equation}
\hat p^\mu \hat p^\nu \to \frac{\vec p^2}{D}(\delta^{\mu\nu}-\delta^{\mu0}\delta^{\nu0}),
\end{equation}
in the integrand. Now, by integrating over $D$ dimensions, we have
\begin{equation}\label{Pi1}
\Pi_{1} = -\frac{8i(\mu^2)^{ \frac{3-D}{2} } }{ (4\pi)^{ \frac D2 } }\,\Gamma\left(2-\frac D2\right)\int\frac{dp_4}{2\pi}m^2(p_4^2+m^2)^{\frac D2-2}
\end{equation}
and
\begin{equation}\label{Pi2}
\Pi_{2} = \frac{4i(\mu^2)^{\frac{3-D}{2}} }{ (4\pi)^{\frac D2} }\,\Gamma\left(1-\frac D2\right)\int\frac{dp_4}{2\pi}\left[(D-1)(p_4^2+m^2)^{\frac D2-1}-(D-2)m^{2}(p_4^2+m^2)^{\frac D2-2}\right].
\end{equation}
Finally, we integrate over the momentum $p_4$ and obtain
\begin{equation}
\Pi_{1} = -\frac{8i(\mu^2)^{\frac{3-D}{2}}}{(4\pi)^{\frac {D+1}2}}\,\Gamma\left(\frac{3-D}2\right)m^{D-1},
\end{equation}
and $\Pi_{2} = 0$.

The analytic expression for the fourth graph of Fig. \ref{fig2}, after transforming to Euclidean space, can be written as
\begin{eqnarray}
\Pi_{bbb}^\mu &=& 4i \int\frac{d^4p_\RM{E}}{(2\pi)^4} \frac{[4(p_\RM{E}^2-m^2)(p_\RM{E}\cdot b_\RM{E})^2-(p_\RM{E}^4-6p_\RM{E}^2m^2+m^4)b_\RM{E}^2]b_\RM{E}^\mu}{(p_\RM{E}^2+m^2)^4}\nonumber\\ 
&&+ 16i \int\frac{d^4p_\RM{E}}{(2\pi)^4} \frac{[(p_\RM{E}^2-m^2)(p_\RM{E}\cdot b_\RM{E})b_\RM{E}^2-2(p_\RM{E}\cdot b_\RM{E})^3]p_\RM{E}^\mu}{(p_\RM{E}^2+m^2)^4}.
\end{eqnarray}
By following the same steps as above for $\Pi_b$ and using also the rotational symmetry of the integrands, which justifies the replacement
\begin{eqnarray}
\hat p^\mu \hat p^\nu \hat p^\lambda \hat p^\rho &\to & \frac{\vec p^4}{D(D+2)}[(\delta^{\mu\nu}-\delta^{\mu0}\delta^{\nu0})(\delta^{\lambda\rho}-\delta^{\lambda0}\delta^{\rho0})+(\delta^{\mu\lambda}-\delta^{\mu0}\delta^{\lambda0})(\delta^{\nu\rho}-\delta^{\nu0}\delta^{\rho0}) \\\nonumber &&+(\delta^{\mu\rho}-\delta^{\mu0}\delta^{\rho0})(\delta^{\lambda\nu}-\delta^{\lambda0}\delta^{\nu0})],
\end{eqnarray}
we obtain
\begin{equation}\label{Pibbb}
\Pi_{bbb}^\mu = \Pi_{3}\,b_\RM{E}^2b_\RM{E}^\mu + \Pi_{4}\, b_4^2b_\RM{E}^\mu + \Pi_{5}\, b_\RM{E}^2b_4\delta^{\mu0} + \Pi_{6}\, b_4^3\delta^{\mu0},
\end{equation}
with
\begin{eqnarray}
\label{Pi3}\Pi_{3} &=& \frac{8i(\mu^2)^{\frac{3-D}{2}}}{3(4\pi)^{\frac D2}}\,\Gamma\left(3-\frac D2\right)\int\frac{dp_4}{2\pi}m^2\left[4(p_4^2+m^2)^{\frac D2-3}+(D-6)m^2(p_4^2+m^2)^{\frac D2-4}\right], \\
\label{Pi4}\Pi_{4} &=& \Pi_5=\frac{8i(\mu^2)^{\frac{3-D}{2}}}{3(4\pi)^{\frac D2}}\,\Gamma\left(3-\frac D2\right)\int\frac{dp_4}{2\pi}m^2\left[(D-5)(p_4^2+m^2)^{\frac D2-3}\right.\nonumber\\&&-\left.(D-6)m^2(p_4^2+m^2)^{\frac D2-4}\right], \\
\label{Pi6}\Pi_{6} &=& -\frac{4i(\mu^2)^{\frac{3-D}{2}}}{3(4\pi)^{\frac D2}}\,\Gamma\left(2-\frac D2\right)\int\frac{dp_4}{2\pi}\left[(D-1)(D-3)(p_4^2+m^2)^{\frac D2-2}\right. \nonumber\\
&&-\left.2(D-3)(D-4)m^2(p_4^2+m^2)^{\frac D2-3}+(D-4)(D-6)m^4(p_4^2+m^2)^{\frac D2-4}\right],
\end{eqnarray}
where we have also integrated over $D$ dimensions. Now, by integrating over the momentum $p_4$, we obtain
\begin{equation}
\Pi_{3} = \frac{8i(\mu^2)^{\frac{3-D}{2}}}{3(4\pi)^{\frac {D+1}2}}\,\Gamma\left(\frac{5-D}2\right)(D-1)m^{D-3},
\end{equation}
and $\Pi_{4} = \Pi_{5} = \Pi_{6} = 0$. By collecting the above results and expanding around $D=3$, the expression for $ \Pi^{\mu} $ can be written as
\begin{equation}\label{PiII}
\Pi^\mu = \left[-\frac{im^2}{\pi^2\epsilon}+\frac{im^2}{2\pi^2}\ln\left(\frac{m^2}{\mu'^2}\right)+\frac{ib_\RM{E}^2}{3\pi^2}\right]b_\RM{E}^\mu,
\end{equation}
with $\epsilon = 3-D$, $\mu'^2=4\pi\mu^2e^{-\gamma}$. Then, after returning to Minkowski space, the expression (\ref{DVef}) can be written as 
\be\label{DVef2}
\left[-\frac1G_\mathrm{R}+\frac{m^2}{2\pi^2}\ln\left(\frac{m^2}{\mu'^2}\right)-\frac{b^2}{3\pi^2}\right]b_\mu = 0,
\en
where we have introduced the renormalized coupling constant
\be
\frac1G_\mathrm{R} = \frac1G + \frac{m^2}{\pi^2\epsilon}.
\en
Therefore, we see that a nontrivial solution of this gap equation is the usual one \cite{Gom}
\be\label{min}
b^2 = -3\pi^2\left[\frac1{G_\mathrm{R}} - \frac{m^2}{2\pi^2}\ln\left(\frac{m^2}{\mu'^2}\right)\right].
\en
From this equation we see that a nontrivial minimum in the case of a timelike $b_\mu$ is possible if
\be\label{cond1}
G_\RM{R} > \frac{2\pi^2}{m^2\ln\left(\frac{m^2}{\mu'^2}\right)},
\en
whereas  a nonzero spacelike $b_\mu$  requires
\be\label{cond2}
G_\RM{R} < \frac{2\pi^2}{m^2\ln\left(\frac{m^2}{\mu'^2}\right)}.
\en

\section{Lorentz symmetry restoration at finite temperature}\label{LSR}

In order to estimate the critical temperature at which restoration of Lorentz and CPT symmetries occurs, we assume that the
 system is in thermal equilibrium at temperature $T$. We use the Matsubara formalism,  which consist in
 taking $p_4=(n+\frac 12)2\pi T$ and replacing $\int\frac{dp_4}{2\pi}=T\sum_n$ in Eqs. (\ref{Pi1}), (\ref{Pi2}), (\ref{Pi3}), (\ref{Pi4}) and (\ref{Pi6}).

\begin{figure}[h]
\includegraphics[scale=0.6]{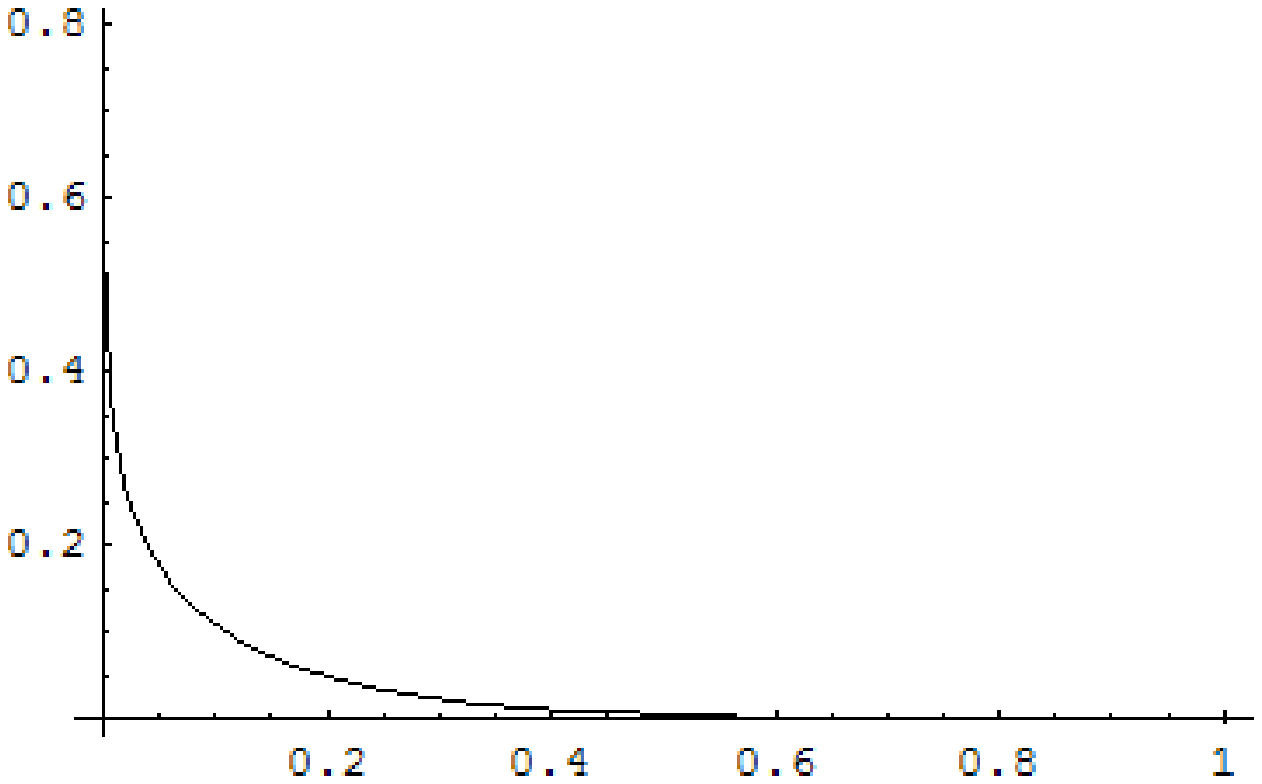}
\caption{Plot of the function $F_\RM{I}(\xi)$} \label{F2}
\end{figure}

\begin{figure}[h]
\includegraphics[scale=0.6]{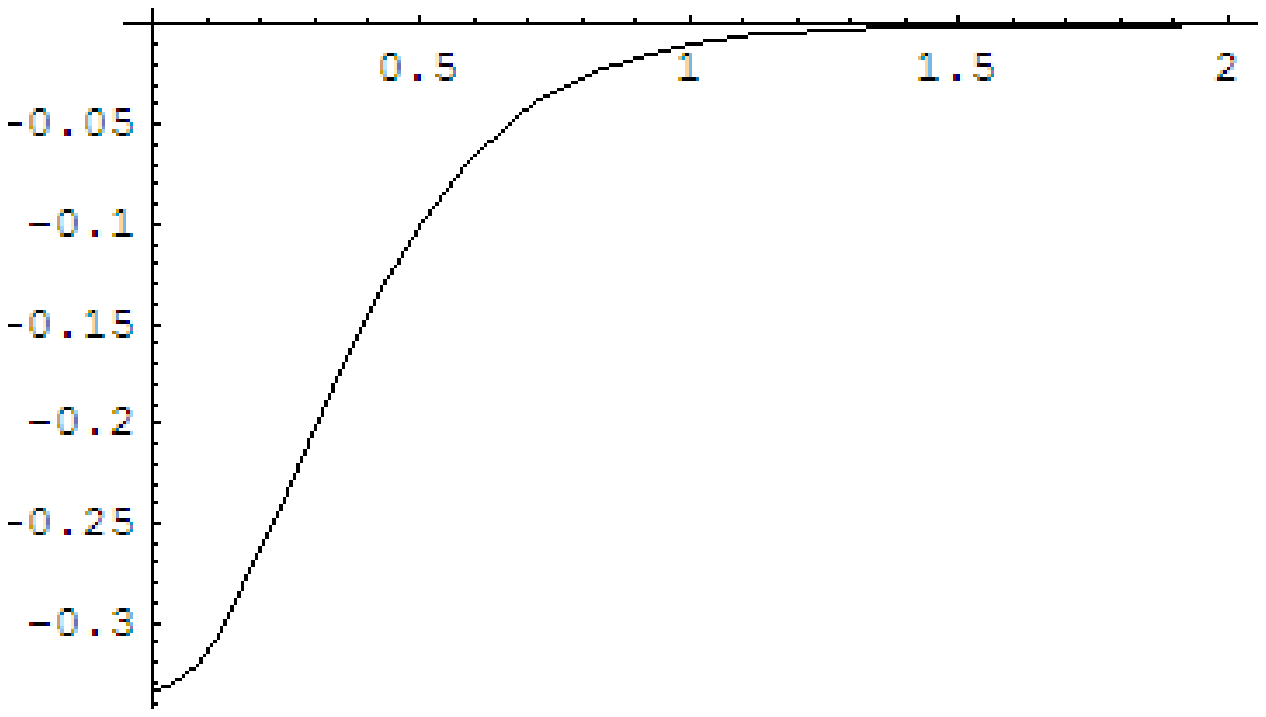}
\caption{Plot of the function $F_\RM{II}(\xi)$} \label{F04}
\end{figure}

Again, we will calculate the momentum integral by adopting the dimensional regularization scheme. To perform the summation we use an explicit representation for the sum over the Matsubara frequencies \cite{Ford}:
\begin{equation}\label{sum}
\sum_n\bigl[(n+b)^2+a^2\bigl]^{-\lambda} = \frac{\sqrt{\pi}\Gamma(\lambda-1/2)}{\Gamma(\lambda)(a^2)^{\lambda-1/2}}+4\sin(\pi\lambda)f_\lambda(a,b)
\end{equation}
where
\begin{equation}
f_\lambda(a,b) = \int^{\infty}_{|a|}\frac{dz}{(z^2-a^2)^{\lambda}}Re\Biggl(\frac{1}{e^{2\pi(z+ib)}-1}\Biggl),
\end{equation}
which is valid for $\lambda<1$ and possesses poles at $\lambda=1/2,-1/2,-3/2,\cdots$. If we take $D=3$ in Eqs. (\ref{Pi1}) and (\ref{Pi2}) we can see that $\lambda<1$, but at $\lambda=1/2,-1/2$ these expressions are singular. Taking the limit $D\to 3$ we find that the  Eq. (\ref{Pib}) becomes
\begin{equation}\label{Pib20}
\Pi_b^\mu = \left[-\frac{im^2}{\pi^2\epsilon}+\frac{im^2}{2\pi^2}\ln\left(\frac{m^2}{\mu'^2}\right)+im^2F_\RM{I}(\xi)\right]b_\RM{E}^\mu + iT^2F_\RM{II}(\xi)b_4\delta^{\mu0},
\end{equation}
where $\xi=\frac{ m}{2\pi T}$ and the functions $F_\RM{I}(\xi)$ and $F_\RM{II}(\xi)$ are given by
\begin{eqnarray}
F_\RM{I}(\xi)&=&\int_{|\xi|}^{\infty}dz\frac{1-\tanh(\pi z)}{\pi^2\sqrt{z^2-\xi^2}},\\
F_\RM{II}(\xi)&=&-\int_{|\xi|}^{\infty}dz\frac{4(\xi^2-2z^2)[\tanh(\pi z)-1]}{\sqrt{z^2-\xi^2}}.
\end{eqnarray}
The plots of these functions are presented in the Figs.~\ref{F2} and \ref{F04}.  Their asymptotic behaviors at high temperature are $F_\RM{I}(\xi\to0)\approx1/2$ and $F_\RM{II}(\xi\to0)=-1/3$ and they vanish at zero temperature. 

\begin{figure}[h]
\includegraphics[scale=0.6]{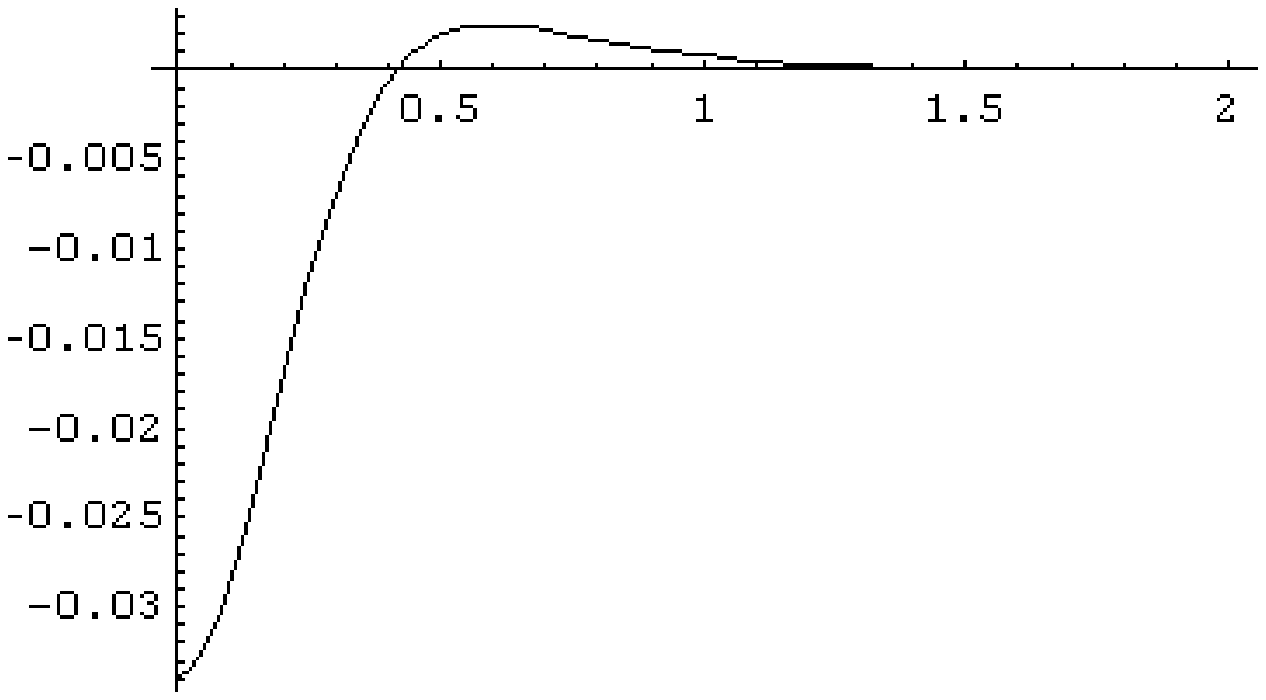}
\caption{Plot of the function $F_\RM{III}(\xi)$}\label{F3}
\end{figure}
\begin{figure}[h]
\includegraphics[scale=0.6]{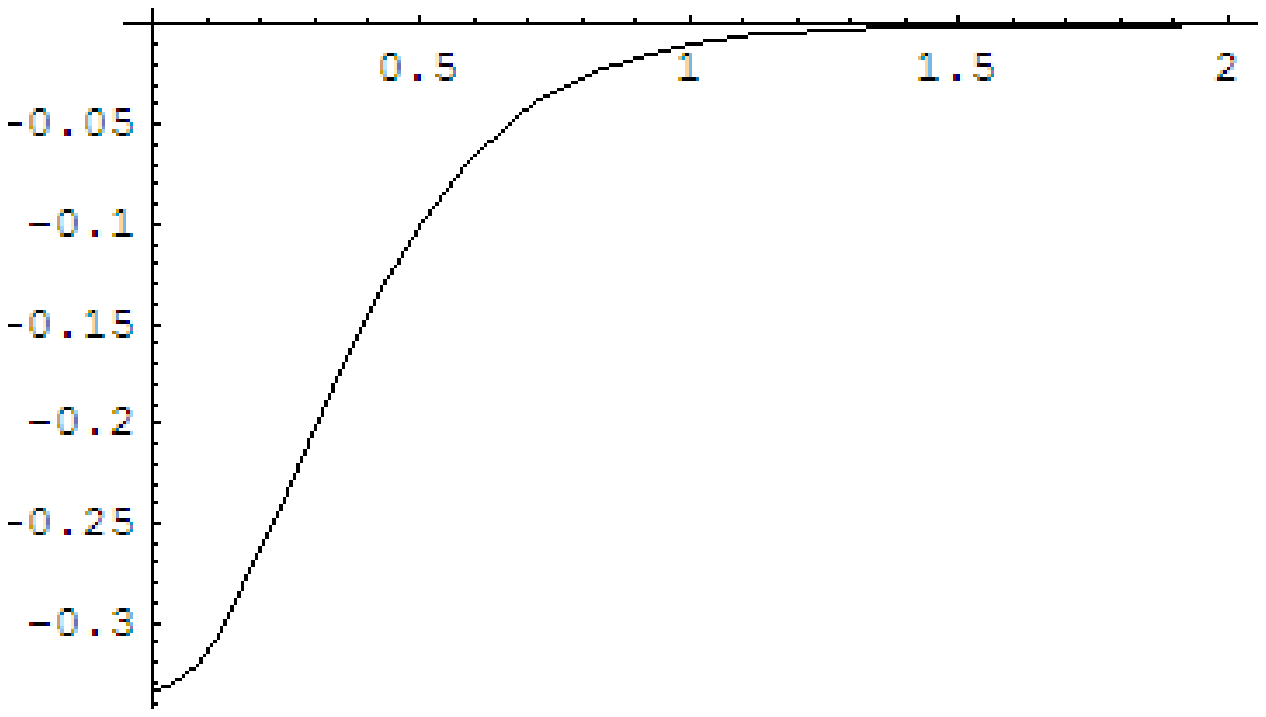}
\caption{Plot of the function $F_\RM{III}(\xi)+F_\RM{IV}(\xi)$}\label{F4}
\end{figure}

By setting $D=3$ in Eqs. (\ref{Pi3}), (\ref{Pi4}), and (\ref{Pi6}), we have $\lambda>1$ and thus we should use the recurrence relation
\begin{equation}\label{rc2}
f_{\lambda-1}(a,b) = -\frac1{2a}\frac{2\lambda-5}{\lambda-2}f_{\lambda-2}(a,b) - \frac1{4a^2}\frac1{(\lambda-3)(\lambda-2)}\frac{\partial^2}{\partial b^2}f_{\lambda-3}(a,b),
\end{equation}
and therefore we now obtain $\lambda=1/2$ and $\lambda=-1/2$ as required for using Eq. (\ref{sum}). Proceeding in this way, we arrive at the following form of the Eq. (\ref{Pibbb}):
\begin{equation}
\Pi_{bbb}^\mu = \left[\frac{ib_\RM{E}^2}{3\pi^2}+ib_\RM{E}^2F_\RM{III}(\xi)+ib_4^2F_\RM{IV}(\xi)\right]b_\RM{E}^\mu + \left[ib_\RM{E}^2F_\RM{IV}(\xi)-ib_4^2 (F_\RM{III}+2F_\RM{IV})\right]b_4\delta^{\mu0},
\end{equation}
with
\begin{eqnarray}
\label{FIII}F_\RM{III}(\xi)&=&\int_{|\xi|}^{\infty}dz\frac{(3\xi^2-2z^2)\,\mathrm{sech}^{2}(\pi z)\tanh(\pi z)}{3\sqrt{z^2-\xi^2}}, \\
\label{FIV}F_\RM{IV}(\xi)&=&F_\RM{V}(\xi)=-\int_{|\xi|}^{\infty}dz\frac{\xi^2\,\mathrm{sech}^{2}(\pi z)\tanh(\pi z)}{3\sqrt{z^2-\xi^2}}.
\end{eqnarray}
These functions  are ploted  in Figs. (\ref{F3}) and (\ref{F4}). At high temperature, their asymptotic behavior are given by $F_\RM{III}(\xi\to0)=-1/3\pi^2$, $F_\RM{IV}(\xi\to0)=0$.

The gap equation (\ref{DVef2}) becomes
\begin{eqnarray}
&&\left[-\frac{1}{G'_\mathrm{R}}+\frac{\vec{b}^2+b^2_4}{3\pi^2}+m^2F_\RM{I}(\xi)+T^2F_\RM{II}(\xi)+\vec b^2\left(F_\RM{III}(\xi)+F_\RM{IV}(\xi)\right)\right]b_4\delta^{\mu0} +\nonumber\\
&&\left[-\frac{1}{G'_\mathrm{R}}+\frac{\vec{b}^2+b^2_4}{3\pi^2}+m^2F_\RM{I}(\xi)+\vec{b}^2F_\RM{III}(\xi)+b^2_4\left(F_\RM{III}(\xi)+F_\RM{IV}(\xi)\right)\right]\hat b^\mu =0,
\end{eqnarray}
where $$\frac1{G'_\RM{R}} = \frac1{G_\RM{R}} - \frac{m^2}{2\pi^2}\ln\left(\frac{m^2}{\mu'^2}\right).$$
Analysing this equation, we find that there are the following possible situations:

(i) $b_4=b_i=0$ is a trivial solution where the Lorentz symmetry is not broken.

(ii) $b_4\neq 0$ and $b_i\neq 0$. In this case $b_4$ and/or $b_i$ quadratically grows with the temperature, so, the Lorentz symmetry breaking increases.

(iii) $b_4\neq 0$ and $b_i=0$. In this case, the gap equation is reduced to
\begin{equation}\label{gapT}
-b_4^2 = b_0^2 = 3\pi^2\left(\frac1{|G'_\RM{R}|}-T^2|F_\RM{II}(\xi)|+m^2F_\RM{I}(\xi)\right),
\end{equation}
where we have used $G'_\RM{R}=-|G'_\RM{R}|$, because $b_\mu$ is
timelike (see Eq. (\ref{cond1})) , and both $F_\RM{I}(\xi)$ and $|F_\RM{II}(\xi)|$
monotonically grow with  $T$. In particular, as $T\to\infty$, $F_I\to
  1/2$ and $|F_{II}|\to 1/3$. Therefore, at a certain temperature, that is, the critical temperature, the condition $b_4=0$ is satisfied.  For  higher temperatures, the term in parentheses in Eq. (\ref{gapT}) 
becomes negative and therefore at temperatures larger than this critical temperature, the Lorentz symmetry will be restored. 

(iv) $b_4= 0$ and $b_i\neq 0$. In this case  the gap equation is
\begin{eqnarray}
-\frac{1}{G'_\mathrm{R}}+m^2F_\RM{I}(\xi)+\vec{b}^2\left(\frac{1}{3\pi^2}+F_\RM{III}(\xi)\right)=0.\label{iv}
\end{eqnarray}
Hence, as $ \frac{1}{3\pi^2}+F_\RM{III}(\xi)\geq 0 $  and $ F_\RM{I}\leq 1/2 $ there will be a critical temperature on which symmetry restoration occurs only if $\frac{1}{m^{2}G'_\mathrm{R}} \leq 1/2 $.  On the other side, for $\frac{1}{m^{2}G'_\mathrm{R}} > 1/2 $, the Lorentz symmetry breaking grows with the temperature.

\section{Summary}\label{Con}

In this paper we have shown that the $4$D four-fermion model, which presents  spontaneous breaking of Lorentz and CPT symmetry at zero temperature, in the high temperature limit can display a critical temperature, 
beyond which the symmetry is restored. Thus, at this temperature a phase transition occurs in the model. However, if $\frac{1}{m^{2}G'_\mathrm{R}} \geq 1/2 $ Lorentz symmetry is broken by the appearance of a space-like vector $ b^{\mu} $ and no restoration occurs at any temperature. 

We would like to compare our results with those ones in Ref. \cite{Grig} where the structure of the radiatively induced Lorentz and CPT violating term was also considered. The analysis in that reference was based in the vector field vacuum polarization tensor $ \Pi_{\mu\nu} $  computed up to linear terms in $ b_{\mu} $ whereas we examined the one point function $ \Pi_{\mu} $ up to third order in that parameter. Notice that in \cite{Grig} $ b_{4} $ is taken to be zero and, as a consequence of their analysis,  the breaking increases with the temperature. In our case the parameter $ b_{\mu} $  is generated  by the vacuum expectation value of the auxiliary vector field used to eliminate a quartic fermionic self-interaction. When our $ b_4 $ is zero two possibilities arisen as discussed in the previous Section; in one of them there is a critical temperature above which the symmetry is restored but in the other possibility the breaking increases with the temperature similarly to the result of \cite{Grig}.

{\bf Acknowledgments.}
This work was partially supported by Funda\c c\~ao de Amparo \`a
Pesquisa do  Estado de S\~ao Paulo (FAPESP), Conselho Nacional de
Desenvolvimento Cient\'\i fico e Tecnol\'ogico (CNPq) and Coordena\c c\~ao de Aperfei\c coamento de Pessoal de N\'\i vel Superior (CAPES: AUX-PE-PROCAD 579/2008).



\begin{thebibliography}{99}
\bibitem{Kostel} V. A. Kostelecky, N. Russell, ``Data Tables for Lorentz and CPT Violation", arXiv: 0801.0287 [hep-ph].
\bibitem{Kos01}V.~A.~Kostelecky and S.~Samuel, Phys. Rev. D {\bf 39}, 683 (1989); V.~A.~Kostelecky and R.~Potting, Nucl. Phys. B {\bf 359}, 545 (1991).
\bibitem{Kost} CPT and Lorentz Symmetry III (ed.: V. A. Kostelecky), World Scientific, Singapore, 2005.
\bibitem{Jac}S.~M.~Carroll, G.~B.~Field and R.~Jackiw, Phys. Rev. D {\bf 41}, 1231 (1990).
\bibitem{Bazeia}D.~Bazeia, T.~Mariz, J.~R.~Nascimento, E.~Passos and R.~F.~Ribeiro, J.~Phys. A {\bf 36}, 4937 (2003).
\bibitem{Kos02}D.~Colladay and V.~A.~Kostelecky, Phys. Rev. D {\bf 55}, 6760 (1997); Phys. Rev. D {\bf 58}, 116002 (1998).
\bibitem{Blu01}R.~Bluhm, ``Overview of the SME: Implications and phenomenology of Lorentz violation'', hep-ph/0506054. 
\bibitem{Coll}D.~Colladay, AIP Conf. Proc. {\bf 672}, 65 (2003) [hep-ph/0301223].
\bibitem{Kos03}V.~A.~Kostelecky and R.~Lehnert, Phys. Rev. D {\bf 63}, 065008 (2001) [hep-th/0012060].
\bibitem{Alts}B.~Altschul and V.~A.~Kostelecky, Phys. Lett. B {\bf 628}, 106 (2005) [hep-th/0509068].
\bibitem{Bert}O.~Bertolami and J.~Paramos, Phys. Rev. D {\bf 72}, 044001 (2005) [hep-th/0504215].
\bibitem{Blu02}R.~Bluhm, ``Effects of spontaneous Lorentz violation in gravity'', arXiv: 0801.0141 [gr-qc]. 
\bibitem{Blu03}R.~Bluhm, Int. J. Mod. Phys. D {\bf 16}, 2357 (2008), hep-th/0607127. 
\bibitem{Blu04}R.~Bluhm and V.~A.~Kostelecky, Phys. Rev. D {\bf 71}, 065008 (2005) [hep-th/0412320]. 
\bibitem{Blu05}R.~Bluhm, S.~H.~Fung and V.~A.~Kostelecky,  Phys. Rev. D {\bf 77}, 065020 (2008), arXiv: 0712.4119 [hep-th]. 
\bibitem{Blu06}R.~Bluhm, N.~L.~Gagne, R.~Potting and A.~Vrublevskis, ``Constraints and stability in vector theories with spontaneous Lorentz violation'', arXiv: 0802.4071 [hep-th].
\bibitem{Gom}M.~Gomes, T.~Mariz, J.~R.~Nascimento and A.~J.~da Silva, Phys. Rev. D {\bf 77}, 105002 (2008), arXiv: 0709.2904 [hep-th].
\bibitem{Grig}L.~Cervi, L.~Griguolo and D.~Seminara, Phys. Rev. D {\bf 64}, 105003 (2001).
\bibitem{Coo} F.~Cooper and J.~P.~Mercader, Phys. Rev. D {\bf 43}, 4129 (1991).
\bibitem{Ford}L.~H.~Ford, Phys. Rev. D {\bf 21}, 933 (1980).
\end{thebibliography}
\end{document}